\documentclass[prc,aps,final,twocolumn,showpacs,showkeys,nofootinbib]{revtex4-1}

\usepackage{graphicx}
\usepackage{mathrsfs}
\usepackage{bm}
\usepackage{color,graphicx}
\usepackage{graphicx,colordvi}

\newcommand{\bel}[1]{\begin{equation}\label{#1}}

\def\be{\begin{equation}}
\def\ee{\end{equation}}
\def\bea{\begin{eqnarray}}
\def\eea{\end{eqnarray}}
\def\l{\label}

\def\p{{\bf p}}

\def\r{{\bf r}}
\def\q{{\bf q}}
\def\k{{\bf k}}

\def\u{{\bf u}}

\def\b{\beta}
\def\om{\omega}
\def\Om{\Omega}

\def\ms{\medskip}
\def\siml{\;\hbox{\kern.1em \lower.7ex \hbox{$\sim$} \kern-1.12em
 \raise.5ex \hbox{$<$} \kern.1em}}
\def\simg{\;\hbox{\kern.1em \lower.7ex \hbox{$\sim$} \kern-1.12em
 \raise.5ex \hbox{$>$} \kern.1em}}

\def\d{\hbox{d}}

\def\l{\label}

\def\p{{\bf p}}

\def\om{\omega}
\def\Om{\Omega}

\def\erf{\mathop{\rm erf}\nolimits}

\def\d{\hbox{d}}

\def\siml{\hbox{\kern.1em \lower.6ex \hbox{$\sim$} \kern-1.12em
          \raise.6ex \hbox{$<$} \kern.1em }}
\def\simg{\hbox{\kern.1em \lower.6ex \hbox{$\sim$} \kern-1.12em
          \raise.6ex \hbox{$>$} \kern.1em }}

\begin{document}

\title{Shear viscosity of nuclear matter}

\author{ A.\ G.\ Magner}
\email{Email: magner@kinr.kiev.ua}
\affiliation{Institute for Nuclear Research NASU, 03680 Kiev, Ukraine}
\author{M.\ I.\ Gorenstein}
\affiliation{Bogolyubov Institute for Theoretical Physics NASU, 03680,
Kiev, Ukraine}
\author{U.\ V.\ Grygoriev}
\affiliation{Taras Shevchenko National University, 03022 Kiev, Ukraine}
\author{ V.\ A.\ Plujko}
\affiliation{Taras Shevchenko National University, 03022 Kiev, Ukraine}

\bigskip

\date{October, 9th 2016}
\ms

\begin{abstract}
Shear viscosity $\eta$
is calculated for the nuclear matter described as
a system of interacting nucleons with the van der Waals (VDW) equation
of state. The Boltzmann-Vlasov
kinetic equation is solved
in terms of the plane waves of the
collective overdamped motion.
In the frequent-collision regime,
the  shear viscosity
depends on the particle-number density  $n$ through the mean-field
parameter $a$, which describes attractive forces in the VDW equation.
In the temperature region $T=15 - 40$~MeV,
a ratio of the shear viscosity to the entropy
density $s$ is smaller than 1 at the
nucleon number density
$n =(0.5 - 1.5)\,n^{}_0$,  where $n^{}_0=0.16\,$fm$^{-3}$ is
the particle density of equilibrium nuclear matter at zero temperature.
A minimum of the $\eta/s$ ratio takes place somewhere in a
vicinity of the
 critical point of the VDW system.
Large values of  $\eta/s\gg 1$ are, however, found in both the
low-density, $n\ll n^{}_0$,
and high-density, $n>2n^{}_0$, regions.
This makes the ideal hydrodynamic approach inapplicable for these
densities.
\end{abstract}



\maketitle

\section{INTRODUCTION}

The shear viscosity $\eta$ and its ratio to
the entropy density $s$  became recently
attractive
(see, e.g., Refs.\
\cite{csernai,shaefer2,wiranata-prc-12,kapusta} 
and references therein)
in connection with a development
of the  hydrodynamic approach to the
relativistic nucleus-nucleus
collisions. Chapman and Enskog (CE) obtained
\cite{chapman,uhlenbeck,huang,silin,fertziger,LLv10}
the shear viscosity $\eta$ in a gas of
non-relativistic particles by using
the Boltzmann kinetic equation (BKE)
for the phase-space distribution function  $f(\r,\p,t) $,
where $\r$ and $\p$ are the particle coordinate and momentum,
respectively, and $t$ denotes the time variable.

The BKE was solved within
the frequent-collision (FC)
regime for which one can use  a perturbation
expansion in a small parameter, e.g.,
$\omega/\nu$, where $\nu$ is the collision frequency and $\omega$
measures the characteristic dynamical variations of the
distribution function $\delta f(\r,\p,t)$.  
In this case
the Boltzmann integral collision term is dominant
as compared to other collisionless terms. For their calculations
the local-equilibrium distribution function $f_{\rm l.e.}$ 
was used 
in a standard form in terms of  
the evolution of particle-number density $n(\r,t)$,
temperature $T(\r,t)$, and collective velocity $\u(\r,t)$. 
The hydrodynamic variables  
$n(\r,t)$ and $\u(\r,t)$
are defined as the zero and first moments of the distribution
function in the momentum space. 
Thus, the  evolution derivative of the distribution
function, $\d f/\d t$, as 
one of the local equilibrium 
distribution,
$\d f_{\rm l.e.}/\d t$, in solving the BKE at the first order in $\omega/\nu $
can be decomposed into terms
proportional to that of $n(\r,t) $, $T(\r,t)$, and 
$\u(\r,t)$.
Using then  the standard closed system of
hydrodynamical equations and condition $\delta T=0$,
one obtains \cite{chapman,silin,fertziger,LLv10}
the expression for the shear  viscosity $\eta$.
For a gas of elastic scattering balls with the diameter $d$,
at the first approximation in
$\omega/\nu$, one finds
\cite{chapman}
\be\l{etaCE}
\eta^{}_{\rm CE}=\frac{5}{16 \sqrt{\pi}}\;\frac{\sqrt{mT}}{d^2}\;,
\ee
where $m$ is the particle mass. The shear viscosity $\eta^{}_{\rm CE}$
appears to be independent of
the particle-number density $n$ because, at the first order in
small parameter $\omega/\nu$, the attractive interaction on 
large distances between particles was neglected to simplify the CE viscosity 
calculations.

Extensions of the  CE method to the relativistic
high energy-density problems are given in
Refs.\ \cite{prakash-pr93,wiranata-prc-12}.
In particular, Eq. (\ref{etaCE}) was reproduced in the
nonrelativistic limit within the CE approach in Ref.\
\cite{wiranata-prc-12}. The mixture of different hadron species
was considered in Ref.\ \cite{gorenstein}.
Several investigations
were devoted to go beyond the
hydrodynamical approach \cite{chapman};
see, e.g.,
Refs.~\cite{abrkhal,brooksyk,balescu,baympeth,kolmagpl,magkohofsh,pethsmith2002,kolomietz1996,kolomietz1998,plujko1999,plujko2001a,plujko2001b,kolomietz2004,spiegel_BVKE-visc_2003,smith2005,review}. In contrast to the CE approach, 
the main problem solved in these works
 was to take into account a self-consistent mean field in calculations 
of the viscosity of Fermi liquids within the Landau quasiparticle theory.

In the present paper, we use  the Boltzmann-Vlasov kinetic equation
(BVKE) for  a system of
interacting nucleons with the van der Waals (VDW)
equation of state.
Therefore, both scattering of particles owing to the hard-core repulsions
and Vlasov self-consistent mean field, owing to the VDW attractive
interaction, are taken into account
in solving the BVKE.

In our consideration, the small dynamical variations $\delta f$ are found
from a linearized BVKE in the simplified form,
\be\l{dfge}
\delta f(\r,\p, t) = f(\r,\p, t)- f^{}_{0}(p)\;,
\ee
where $f^{}_{0}(p)$ is the static global-equilibrium
distribution function
\be\l{maxwell}
  f^{}_{0}(p)=
\frac{n}{(2m\pi T)^{3/2}}\; \exp\left(-~\frac{p^2}{2mT}\right)\;.
\ee
The function $f^{}_{0}(p)$ (\ref{maxwell}) 
is taken in the Maxwell form 
with constant
values of $T$ and $n$ and zero value of the collective velocity, $\u=0$. 
The damping plane wave (DPW) solutions are assumed to be a good
approximation to the dynamical variations $\delta f$  
at finite
frequencies $\omega$ within the FC regime (large-enough collision
frequency $\nu$).
These dynamical deviations 
allow us
to take into account analytically the attractive VDW interactions
through the
self-consistent Vlasov mean field in the BVKE. 
As a result, we obtain  the shear viscosity dependence on the
particle number density at the leading  
order in a small parameter $\omega/\tau$. The
overdamped  (see, e.g., Refs.\ \cite{magkohofsh,review}) attenuation
of the DPW will be considered. 
Our approach is
based on the  methods applied earlier for calculations of the viscosity
of the Fermi liquids \cite{abrkhal,brooksyk,baympeth,kolmagpl,review}.
In the present paper, the shear viscosity in the first  order over 
small parameter $\omega/\nu$
is calculated analytically for
the
nuclear matter considered as a gas of interacting
nucleons with the VDW equation of state.

The paper is organized as follows. In Sec.\ \ref{sec-vdw}
we remind the basic  properties
of thermodynamically equilibrated systems
with the VDW equation  of state.
In Sec.\ \ref{sec-kinetic} we outlook
the kinetic approach based on the BVKE and
give general definitions of the  shear viscosity coefficient.
In Sec.\ \ref{sec-disp}, the solution to the BVKE
and its perturbation expansion is  presented
in terms of the
plane waves accounting for a strong attenuation owing
to the particle collisions.
Finally, this section is devoted to
the main results for the VDW viscosity.
The obtained  results are discussed
in Sec.\ \ref{sec-disc} and summarized
in Sec.\ \ref{sec-concl}. Some details of our calculations can be found
in Appendixes A--C.

\section{VDW EQUATION OF STATE}
\label{sec-vdw}

The VDW
equation of state presents the system pressure $P$
in terms of the particle number density $n$
and temperature $T$
as \cite{LL},
\be\l{vdw-p-n}
P(T,n)~=
\frac{n\,T}{1-bn}~-~a\,n^2~,
\ee
where $a>0$ and $b>0$ are the VDW parameters that describe
attractive and repulsive interactions, respectively.
The first term on the right-hand side of Eq.~(\ref{vdw-p-n})
contains the excluded volume correction
($b=2 \pi d^3/3$, with $d$ being the particle hard-core diameter),
while the second term comes from the mean-field description of
attractive interactions.

The entropy density $s$ and energy density $\varepsilon$
for the VDW system are calculated
as \cite{LL}
\bea\label{s}
\!s(T,n)
 &\!=\!& \frac{5}{2}n +
n\ln\!\left[\frac{(1-bn)}{n} g
\left(\frac{mT}{2\pi}\right)^{3/2}\!\right],\\
\varepsilon(T,n)&\!=\!& n \left[\frac{3}{2}T - a\,n\right]\,.\label{E}
\eea
In Eq.~(\ref{s}) $m$ is the particle mass and $g$ is the degeneracy factor
($g=4$ for nucleons; two spin and two isospin states).
Note that
the VDW entropy density (\ref{s})
is independent of the attractive
mean-field interaction parameter $a$, whereas the  energy density (\ref{E})
does not
depend on the particle repulsion constant $b$.

The VDW equation of state contains the first-order liquid-gas phase
transition with a critical point
\cite{LL}:
\be\label{Tc}
 T_c = \frac{8a}{27b}~,~~~~ n_c =
\frac{1}{3b}~,~~~~ P_c = \frac{a}{27b^2}~.
\ee
To study the phase coexistence
region which exists below the critical temperature, $T<T_c$, the VDW isotherms
should be corrected by the well-known Maxwell construction of equal areas.

The VDW equation of state was recently applied to a description of nuclear matter in
Ref.~\cite{marik2}.
In the present study we fix the VDW parameters for the system of interacting nucleons
as $d=1$~fm, i.e., $b\cong 2.1$~fm$^{3}$, and $a=100$~MeV\,fm$^{3}$.
This gives $n_c\cong n_0=0.16$~fm$^{-3}$ and $T_c\cong 14$~MeV
 ($n_0=0.16$~fm$^{-3}$
corresponds to the nucleon number density of the normal
nuclear matter at zero temperature).
In what follows we restrict our analysis of the kinetic properties of the VDW system of nucleons
to $T>T_c$. In this region of the phase diagram the VDW equation of
state describes
a homogeneous one-phase system,
and all criteria of the thermodynamical stability are satisfied.
We do not consider too large temperatures by taking
$T \siml 40$~MeV.
This allows us to neglect a production of new particles
(pions and baryonic resonances) in the system of interacting nucleons.
In addition, this restriction guarantees a good accuracy
of the nonrelativistic approximation adopted in the present study.
Note also that at $T\rightarrow 0$ the quantum statistics effects
neglected in the present study should
be taken into account (see Ref.~\cite{marik2}).

\section{KINETIC APPROACH}\label{sec-kinetic}
For calculations of the shear viscosity,
we start with 
the BVKE
linearized near the static 
distribution function (\ref{maxwell}) for the dynamical variations
of the distribution function $\delta f(\r,\p,t)$ (\ref{dfge}) : 
\be\l{Boltzlin}
\frac{\partial \delta f}{\partial t}+
\frac{{\bf p}}{m} \frac{\partial \delta f}{\partial {\bf r}} -
\frac{\partial f_{0}}{\partial {\bf p}}\;
\frac{\partial \delta U}{\partial {\bf r}}
= \delta St\;.
\ee
The dynamical part of the attractive potential
$\delta U$ from
the VDW forces  is defined self-consistently as
\be\l{effpot}
\delta U(\r,t)=-a\;\int  \d\p\; \delta f(\r,\p,t)\;.
\ee
In Eq.\ (\ref{Boltzlin}), the collision term $\delta St$
is taken in the standard Boltzmann form
\cite{chapman,silin},
\be\l{dstdef}
\delta St~ = ~\frac{2 \pi}{m} \int \d\p_1\; |\p_1-\p| \; \int \b \d \b\;
\delta Q\;,
\ee
where
\bea\l{dQ}
\delta Q& \approx &f^{}_{0}(p')\;\delta f(\r,\p_1',t)+
f^{}_{0}(p_1')\;\delta f(\r,\p',t)\nonumber\\
&-&f^{}_{0}(p)\delta f(\r,\p_1,t)
-f^{}_{0}(p_1)\;\delta f(\r,\p,t)\;
\eea
is the variation of
$f(\r,\p',t) f(\r,\p_1',t)-f(\r,\p,t) f(\r,\p_1,t)$ over $\delta f$ and
$\b$ the impact parameter for two-body collisions.
%
\begin{figure}
\begin{center}
\includegraphics[width=0.49\textwidth,clip=true]{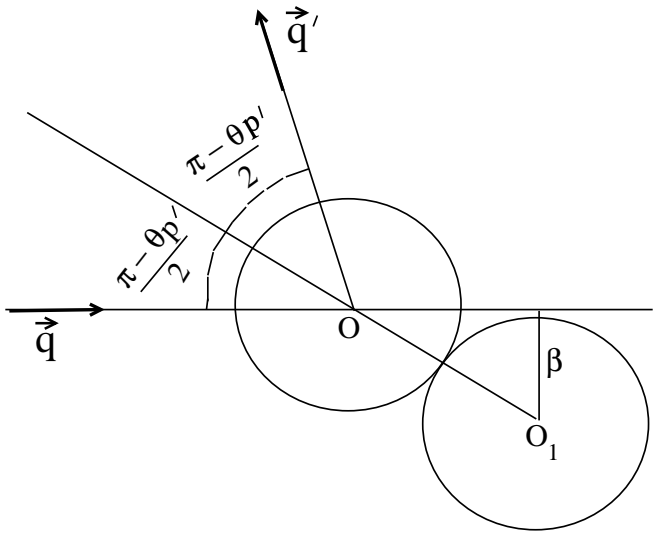}
\end{center}
\caption{{\small The geometry of the collision of
two hard spheres in the center
mass system; $\q=\p_1-\p$, $\q'=\p_1'-\p'$, $OO_1=d$, $d$ is the sphere
particle diameter, $\beta$
is the impact parameter;
$\theta_{p'}$ is the scattering angle.
}}
\label{fig1}
\end{figure}
  Figure \ref{fig1} shows the collision geometry for
two hard-core sphere scattering
in the center-of-mass coordinates. The relationship between
the impact
 parameter, $\beta$, and the cross section, $\sigma=\pi d^2$,
where $d$ is the diameter of the particle, is given in Appendix A
for calculations of the integral term (\ref{dstdef}).

The shear
viscosity $\eta$ can be defined through the
dynamical components of the momentum flux tensor $\Pi_{\mu \nu}(\r,t)$,
\be\l{dpidef}
\delta \Pi_{\mu \nu}
= -\delta \sigma_{\mu \nu} + \delta \mathcal{P}\;\delta_{\mu\nu}
+ \delta P\,\delta_{\mu \nu}\;,
\ee
where $\delta \sigma_{\mu \nu}$ is a traceless stress tensor.
Other terms are diagonal kinetic and interaction pressures.
The stress tensor $\delta \sigma_{\mu \nu}$ can be determined
through the second $\p$ moment of the distribution
function linearized over $\delta f$ as
\be\l{dsigmadef}
\!\delta \sigma_{\mu \nu}=-\int \frac{\d \p}{m}\; p_\mu p_\nu\;
\delta f(\r,\p,t) + \delta \mathcal{P}\;\delta_{\mu\nu}\;.
\ee
In Eq.\ (\ref{dpidef}), the
quantities $\delta \mathcal{P}$ and $\delta P $
are calculated as
\bea\l{presskin}
\delta \mathcal{P} ~& = & ~
\frac{1}{3m} \int p^2\,\d \p\; \delta f(\r,\p,t)\;,\\
\delta P ~ &=& ~ -2 a n \int  \d\p\; \delta f(\r,\p,t)\;.
\eea

 The shear viscosity $\eta$ is defined as a coefficient
in the relationship between
the dynamical component of the stress tensor $\delta \sigma_{\mu \nu}$
[Eq.\ (\ref{dpidef})] and the traceless
tensor $\mathcal{U}_{\mu \nu}$ of the coordinate derivatives of the
velocity field $\u(\r,t)$
\cite{LLv10,balescu,brooksyk,kolmagpl,LLv6},
\be\l{etadef}
\delta \sigma_{\mu\nu}(\r,t)=\eta\; \mathcal{U}_{\mu \nu}\left(\r,t\right)\;,
\ee
where
\be\l{veltens}
\mathcal{U}_{\mu \nu}=
\left(\frac{\partial u_\mu}{\partial r_\nu}
+\frac{\partial u_\nu}{\partial r_\mu}
-\frac{2}{3} \nabla \u \;\delta_{\mu\nu}\right)\;.
\ee
The velocity field $\u$ is defined through
the first $\p$ moment of
$\delta f(\r,\p,t)$,
\be\l{udf}
\u=\frac{1}{n}\;\int \d \p\; \frac{\p}{m}\;\delta f(\r,\p,t)\;.
\ee

Note that our method can also be presented
within the linear response-function theory
\cite{balescu,hofmann,review} (cf. the 
Kubo formulas for the 
diffusion, thermal 
conductivity, and viscosity; see
also the recent article  \cite{wiranata-jp-14}).

\section{DISPERSION RELATION AND VISCOSITY}\label{sec-disp}

We suggest to calculate
the shear viscosity $\eta$
by directly solving the BVKE
(\ref{Boltzlin})
in terms of  the
plane-wave representation for the dynamical
distribution-function variations
$\delta f(\r,\p,t)$ [Eq.\ (\ref{dfge})] in the following
rather general form
 \cite{brooksyk,kolmagpl,review},
\be\l{planewavesol}
\!\delta f(\r,\p,t)\!=\!
f^{}_{0}(p) \varphi(\hat{p})
\exp\left(-i \omega t + i\k \r\right),
\ee
where $\omega$ and $\k$ are a frequency and a wave vector  of the DPW,
respectively.  As unknown yet,  amplitudes
$\varphi(\hat{p})$ are
functions of the momentum angle variable $\hat{p}=\p/p$.
It is naturally to
find solutions of the BVKE as proportional to the static distribution
function, $f^{}_{0}(p)$, specifying the dependence of $\delta f$
on the modulus of
momentum $p$ because the derivative of $f^{}_{0}(p)$ (\ref{maxwell})
over momentum
in Eq.\ (\ref{Boltzlin})
and the variations of the collision integral (\ref{dstdef})
are proportional to  $f^{}_{0}(p)$. Then, one can reduce the problem
for solving the BVKE (\ref{Boltzlin}) to a function of angles
$\varphi(\hat{p})$
which, however, depends on the uknown frequency $\omega~$.
[We shall leave out the argument $\omega$ in $\varphi(\hat{p})$
for simplicity of the notations.] Note that any physical quantity,
in particular the viscosity coefficient,
is independent of  the direction of the unit wave vector $\hat{k}=\k/k$
of the DPW  spreading
in infinite
nuclear matter. Therefore,
it is convenient to  use the spherical phase-space coordinate system
with the polar
axis directed to this vector $\hat{k}$.
The solution for the plane-wave distribution function
$\delta f(\r,\p,t)$ [Eq.\ (\ref{planewavesol})]
or more precisely $\varphi(\hat{p})$, and the frequency
$\omega $ depends only on the wave
vector length $k$. For convenience, one may write the
frequency $\omega$ through the wave number $k$ and
the dimensionless sound velocity $c$,
\be\l{freqom}
\omega= k v =
kv^{}_{T} c\;,
\ee
where
$v=v^{}_{T} c$ is the DPW speed,
and $c$ its dimensionless value
given in units of the most probable thermal
velocity $v^{}_{T}$
of particles at a given temperature $T$, $v^{}_{T}=\sqrt{2T/m}$.

The viscosity $\eta$ is related to an attenuation of the DPW
(\ref{planewavesol})  measured by
the collision term $\delta St$ (\ref{dstdef}).
Following Refs.\ \cite{chapman,silin,fertziger,brooksyk},
one applies the perturbation
expansion  of the dynamical distribution-function variations $\delta f$
through their amplitudes $\varphi(\hat{p})$,
\be\l{varphipertexp}
\varphi(\hat{p})=\varphi^{(0)}(\hat{p}) + \epsilon \varphi^{(1)}(\hat{p}) +
\epsilon^2 \varphi^{(2)}(\hat{p}) + ...,
\ee
and similarly, in addition to Ref.\ \cite{chapman}, for the
frequency $\omega$,
\be\l{ompertexp}
\omega=\omega^{(0)} + \epsilon \omega^{(1)} +
\epsilon^2 \omega^{(2)} + ...,
\ee
in a small parameter,
\be\l{eps}
\epsilon=\omega/\nu =\omega \tau\;. 
\ee
Here, $\tau$ is the relaxation time\footnote{We do no not use the
standard $\tau$ approximation and introduce
the relaxation time $\tau$  for sake of the convenience in comparison
with other approaches.}  defined by the
collision term through the time-dependent
rate $\nu$ (collision frequency) of the damping of
distribution function $\delta f$,
\be\l{taudef}
\tau=1/\nu\;.
\ee
Expansions (\ref{varphipertexp}) and (\ref{ompertexp}) are
defined within the standard perturbation method
\cite{silin,brooksyk,review,madelung}
for the eigenfunction,
$\varphi(\hat{p})$, and
eigenvalue, $\omega$, problem.
In these perturbation expansions,
the coefficients $\varphi^{(n)}(\hat{p})$ and $\omega^{(n)}$
are assumed to be
independent of $\epsilon $.  By using the BVKE with this
perturbation
method, they can be
found at each order of $\epsilon$.
Note that $\omega$ in the definition of the  small
parameter $\epsilon$ (\ref{eps}) is determined consistently at any
given order of the perturbation expansions
(\ref{varphipertexp}) and (\ref{ompertexp}); see Appendix B for details.
A smallness of $\epsilon$ 
can be achieved by increasing  
the collision frequency  $\nu$ 
for a given
$\omega$ (Appendix A).
Substituting the plane-wave representation
(\ref{planewavesol}) for the distribution function $\delta f$ into the BVKE
(\ref{Boltzlin}), for convenience, one can also
expand $\varphi(\hat{p})$ in series
over the spherical harmonics $Y_{\ell 0}(\hat{p})$, %
\be\l{varphiexp}
\varphi(\hat{p})=\sum_{\ell=0}^{\infty}
\varphi^{}_{\ell}\;Y_{\ell 0}\left(\hat{p}\right)\;,\quad
\hat{p}=\frac{\p \cdot \k}{pk}\;.
\ee
This reduces the integro-differential BVKE to much more simple linear algebraic
equations (\ref{Boltzeqfin}) for the partial multipole amplitudes
$\varphi^{}_{\ell}$ at each order in $\epsilon$  [Eq.\ (\ref{eps}) and
Appendix B].

As shown in Appendixes A and B, in
the FC regime,
$|\epsilon| \ll 1$ [Eq.\ (\ref{eps})], one can  truncate
the multipole expansion (\ref{varphiexp}) over $\ell$ at $\ell=2$ because of
a good convergence in the small  parameter $\epsilon~$.
At this leading
approximation to viscosity calculations, for the collision
term $\delta St$ [Eq.\  (\ref{dstdef})], one obtains (Appendix A)
the simple expression
\be\l{tauapprox}
\delta St = -\nu\; \delta f^{}_2(\r,\p,t)\;,
\ee
where
\be\l{nu}
\nu \approx
\frac{3 n v^{}_{T}\sigma}{2}\;,\qquad \sigma=\pi d^2\;,
\ee
$\sigma$ is the cross section
for a two elastic  hard-core sphere scattering,
as introduced above,
\be\l{df2}
\!\delta f_2(\r,\p,t)\!=\!f^{}_{0}(p)\varphi^{}_2Y_{20}(\hat{p})
\exp\left(-i \omega t + i\k \r\right)\;.
\ee
As shown in Appendix A,
within the accuracy about ~~6\%, this value agrees with its
 mean effective quantity $\nu_{av}$
(\ref{nuav}), evaluated through the
momentum average of the collision term,
$\langle \delta St \rangle_{\rm av}$,
over particle momenta $p$ with the help of the
Maxwell distribution $f^{}_{0}(p)$ (\ref{maxwell}).
Multiplying then the BVKE (\ref{Boltzlin}), with the quadrupole
collisional term
(\ref{tauapprox}), by the spherical function
$Y_{L 0}(\hat{p})$ ($L=0,1,2,...$), one can integrate the BVKE
term by term over angles
($\hat{p}$) of the momentum $\p$. Thus, one obtains the linear
homogeneous equations
(\ref{Boltzeqfin}) with respect to coefficients $\varphi^{}_\ell$
of the expansion
(\ref{varphiexp}) in the plane-wave amplitudes $\varphi(\hat{p})$
at any order in $\epsilon$ in
Eqs.\ (\ref{varphipertexp}) and (\ref{ompertexp}).
This system has nontrivial solutions
in the quadrupole approximation
$\ell \leq 2$,
valid at the
leading  (linear in $\epsilon$) approximation in expansions
(\ref{varphipertexp}) and (\ref{ompertexp}). They obey
the cubic dispersion equation
for $c=\omega/(k v^{}_{T})$ (expansion of $c$ is similar to Eq.\
(\ref{ompertexp}); see also Appendix B),
\bea\l{dispeq}
& {\rm det}\mathcal{A}_2
\equiv
c^3 + i \gamma\;c^2  - c\left[\frac{4}{15} +
\frac{1}{3}\left(
1 - \mathcal{F}\right)\right]\nonumber\\
& - \frac{i}{3}\;
\left(1 - \mathcal{F}\right)\gamma =0\;,
\eea
where $\mathcal{F}$ is the dimensionless VDW interaction parameter,
\be\l{Fant}
\mathcal{F}=an/T\;.
\ee
The truncated (at $\ell=2 $) $3 \times 3$ matrix
$\mathcal{A}^{(2)}_{L\ell}(c)$ is given by Eq.\ (\ref{matrix2}).
For convenience, we introduced also the dimensionless
collisional rate (\ref{nu}):
\be\l{gamma}
\gamma=\frac{\nu}{k v^{}_{T}}=\frac{\nu c}{\omega} =
\frac{c}{\omega \tau}\;.
\ee
The FC perturbation parameter
$\epsilon$ [Eq.\ (\ref{eps})] can be
expressed in terms of the
$\gamma$ and $c$ as
\be\l{FCcond}
\epsilon=c/\gamma\;,\quad |c/\gamma| \ll 1\;.
\ee

The cubic dispersion equation (\ref{dispeq})
has still two limit solutions
with respect to the complex velocity, $c=c_r+ic_i$ for real $k$
(or equivalently, a complex
wave number $k=k_r+ik_i$ for a real velocity $c$, both
related by the same $\omega=k c v^{}_{T}=\omega_r + i \omega_i$, where
low subscripts denote the real and imaginary parts).
One of them can be called
as the underdamped (weakly damped) first sound mode for which
the imaginary part of $c$,
$c_i$, is much smaller than the real one $c_r$, $|c_i/c_r| \ll 1$, while in
the opposite case $|c_i/c_r| \gg 1$, one has the overdamped motion.
In the first
underdamped sound case ($|c_i/c_r| \ll 1 $),
the  collision term can be considered
as small with respect to the 
left-hand side (LHS)
of the BVKE,
$|\gamma/c| \sim 1/|\omega \tau| \ll 1$, that is, the rare collision  
(RC)
regime.
For the overdamped
motion  (FC case) the collision term is 
dominant. In our DPW derivations below one can use also the frequency
expansion
(\ref{ompertexp})  over the same small parameter
$|c/\gamma| \sim |\omega \tau| \ll 1$ [ Eqs.\ (\ref{gamma}) and
(\ref{FCcond})]. In the present study we consider 
the overdamped motion, while the underdamped case
will be studied in 
separate publications.

Expanding
the LHS of the truncated (quadrupole) dispersion equation (\ref{dispeq})
for $c$ in powers of $\epsilon$ [see Eqs.\  (\ref{eps})
and (\ref{FCcond})] in the 
FC perturbation
expansions (\ref{varphipertexp}) and (\ref{ompertexp}),
one can divide all of its terms by $\gamma^3$.
Then, one can  neglect the
relatively small cubic [$(c/\gamma)^3\sim \epsilon^3$] and quadratic
($\sim \epsilon^2$) terms as compared to
the last two linear (in $\epsilon$) ones
depending explicitly on the interaction  parameter
$\mathcal{F}$ [Eq.\ (\ref{Fant})]. At this leading order,
one results in the explicit quadrupole
solution for the velocity $c$ [Eq.\ (\ref{sol0soundr})],
\be\l{linsols0}
c = i c_i
=-\frac{5i}{9}\;\frac{1 - \mathcal{F}}{1 - 5 \mathcal{F}/9}\; \gamma\;.
\ee
To get small corrections of the real sound velocity $c_r$,
one has to take into account the quadratic and cubic in $c$ terms of
the dispersion
equation (\ref{dispeq}).
Note that formally, one can consider
the real DPW velocity $c$ but the complex
wave number $k$ within the same complex
frequency $\omega$, which are both almost pure imaginary ones.
The latter describes the sound attenuation as the
exponential decrease of the
DPW amplitude, $\delta f \propto \exp(- t/\mathcal{T})$
[Eq.\ (\ref{planewavesol})] with the damping time
$\mathcal{T}$,
\be\l{damptime}
\mathcal{T} \approx
\frac{6}{5\pi}\;\frac{1 - 5\mathcal{F}/9}{(1 -
\mathcal{F}) n v^{}_{T}d^2}\;.
\ee
This time was obtained as the imaginary part of the
complex frequency, $\omega =
-i/\mathcal{T}$,  through
Eq.\ (\ref{linsols0}), formally introduced above
(finally, all physical
quantities will be determined by taking their real parts).
Note also that the relaxation time $\tau$ [Eqs.\ (\ref{taudef})
and (\ref{nu})],
\be\l{reltime}
\tau =\frac{1}{\nu}\approx
\frac{2}{3 n v^{}_{T}\pi d^2}\;,
\ee
differs
from the damping time, $\mathcal{T}$ [Eq.\ (\ref{damptime})].
In particular, this time $\mathcal{T}$, being of the order of $\tau$,
depends on the interaction constant $\mathcal{F}$. 
Note that the FC condition
(\ref{FCcond}) can be satisfied for the interaction parameter
$\mathcal{F}$ of the order of one. However, as shown below, one finds
a reasonable result even in the limit $\mathcal{F}\to 0$.

\begin{figure}
\begin{center}
\includegraphics[width=0.49\textwidth,clip=true]{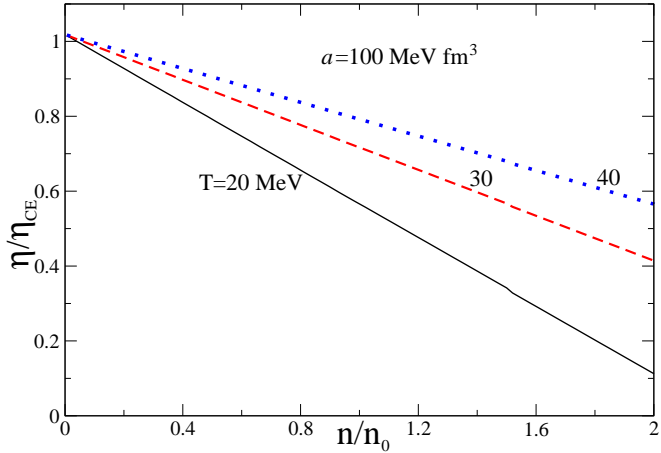}
\end{center}

\vspace{-0.7cm}
\caption{{\small
Shear viscosity $\eta$  [Eq.\ (\ref{viscSDfc})]
in the frequent-collision regime in units of the
CE value $\eta^{}_{\rm CE}$ [Eq.\ (\ref{etaCE})] versus particle-number density
$n$ in units of the normal density $n_0=0.16$ fm$^{-3}$ of nuclear matter;
 $m\cong 938$~MeV;
$d=1~$fm, $a=100$~MeV fm$^3$.
}}
 \label{fig2}
\end{figure}

Using the DPW solutions (\ref{planewavesol}) for $\delta f$
of the BVKE (\ref{Boltzlin}), and
Eqs.\ (\ref{USzz}) for $\mathcal{U}_{zz}$ and (\ref{Szztilde}) for
$\sigma_{zz}$,  for
the definition of the shear viscosity $\eta$ [Eq.\ (\ref{etadef})], one finds
the FC expansions
[(\ref{varphipertexp}) and (\ref{ompertexp})] of $\eta$
 in powers of small $\epsilon$  [see
 Eq.\ (\ref{eps}), and Appendixes C and B].
As shown in Appendix C, the leading term of this FC shear viscosity
$\eta$ at first order in $\epsilon$ is approximately a constant,
independent of $\omega$ (or $k$), and
 proportional to $1/\nu$, i.e., to the relaxation time
$\tau$ [Eq.\ (\ref{reltime})]. Finally, up to relatively high
(second)
order terms in the small parameter $\epsilon$ we arrive at
\bea\l{viscSDfc}
\!\eta \!&=&\! \frac{9}{20 \sqrt{2\pi}}\;
\left(1\!-\!\frac59 \frac{an}{T}\right)\frac{\sqrt{mT}}{d^2}
\!=\!
\frac{36}{25 \sqrt{2}}\;
\left(1\!-\!\frac59 \frac{an}{T}\right)\eta^{}_{\rm CE}\nonumber\\
&=&
1.018\left(1-\frac59 \frac{an}{T}\right)\;\eta^{}_{\rm CE}\;.
\eea
In these derivations we used, at the leading 
first order in $\epsilon$, the
quadrupole multipolarity
truncation of rapidly converged series (\ref{varphiexp});
see Eqs.\ (\ref{phieq}) for the amplitudes $\varphi^{}_\ell$
and (\ref{linsols0})
for the sound velocity $c$
($\ell \leq 2$) within the dispersion
equation (\ref{dispeq}). 
As seen from 
Eq.\ (\ref{viscSDfc}),   
within the present accuracy, the shear
viscosity $\eta$   
differs in  2 \%  
from the CE result $\eta^{}_{\rm CE}$ [Eq.\ (\ref{etaCE})]
at zero attractive mean field, $a \to 0$. 
Note that  
a more exact CE result is 
$\eta=1.016 \eta_{\rm CE}$ (see Ref.\ \cite{chapman}, Chap. 12.1).

Formula (\ref{viscSDfc})
for the shear viscosity
can be presented in a more traditional way through the relaxation time
$\tau$ [Eq.\ (\ref{reltime})],
\be\l{etatau2}
\eta = \frac{27 \sqrt{\pi}}{80}\;\left(1 - \frac{5}{9}
 \mathcal{F}\right)\;
m n v^2_{T}\tau\;.
\ee
This relationship,  $\eta \propto \tau$,
is typical for the 
FC  regime, in contrast
to the rare collision one, $\eta \propto 1/\tau$,
which should be expected for the perturbation expansion at leading order in
the opposite small parameter $1/\epsilon$; see
Refs.\ \cite{brooksyk,baympeth,kolmagpl,review}.
\begin{figure*}
\begin{center}
\includegraphics[width=0.8\textwidth,clip=true]{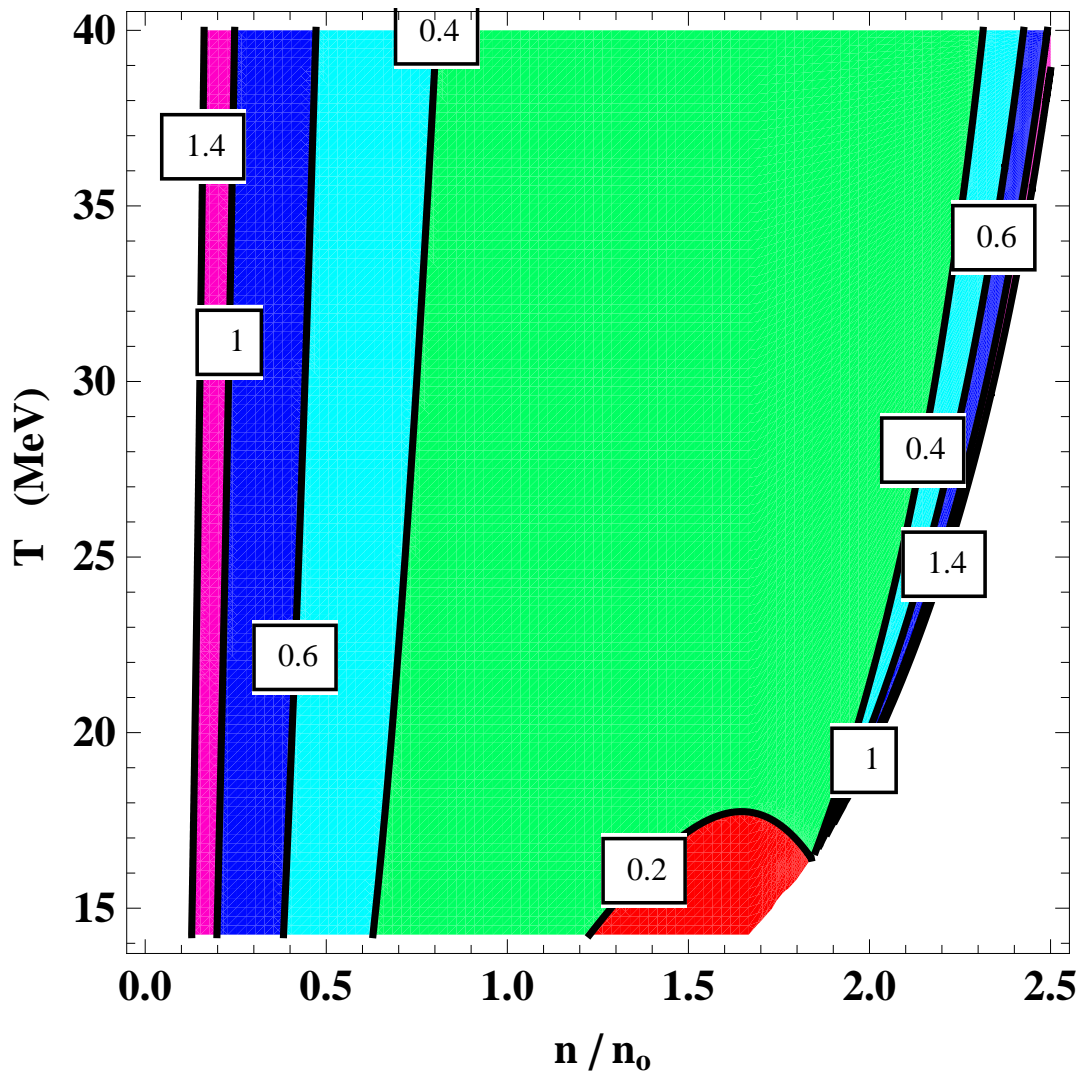}
\end{center}

\vspace{-0.7cm}
\caption{{\small
Contour plot for the ratio  $\eta/s$ of the DPW viscosity
$\eta$ [Eq.\ (\ref{viscSDfc})] to the entropy density
$s$ (\ref{s}) in the $n-T$  plane
at $g=4$ and the same $m$, $d$, and $a$ parameters as in Fig.\ \ref{fig2}.
 }}\label{fig3}
\end{figure*}
\begin{figure*}
\begin{center}
\includegraphics[width=0.8\textwidth,clip=true]{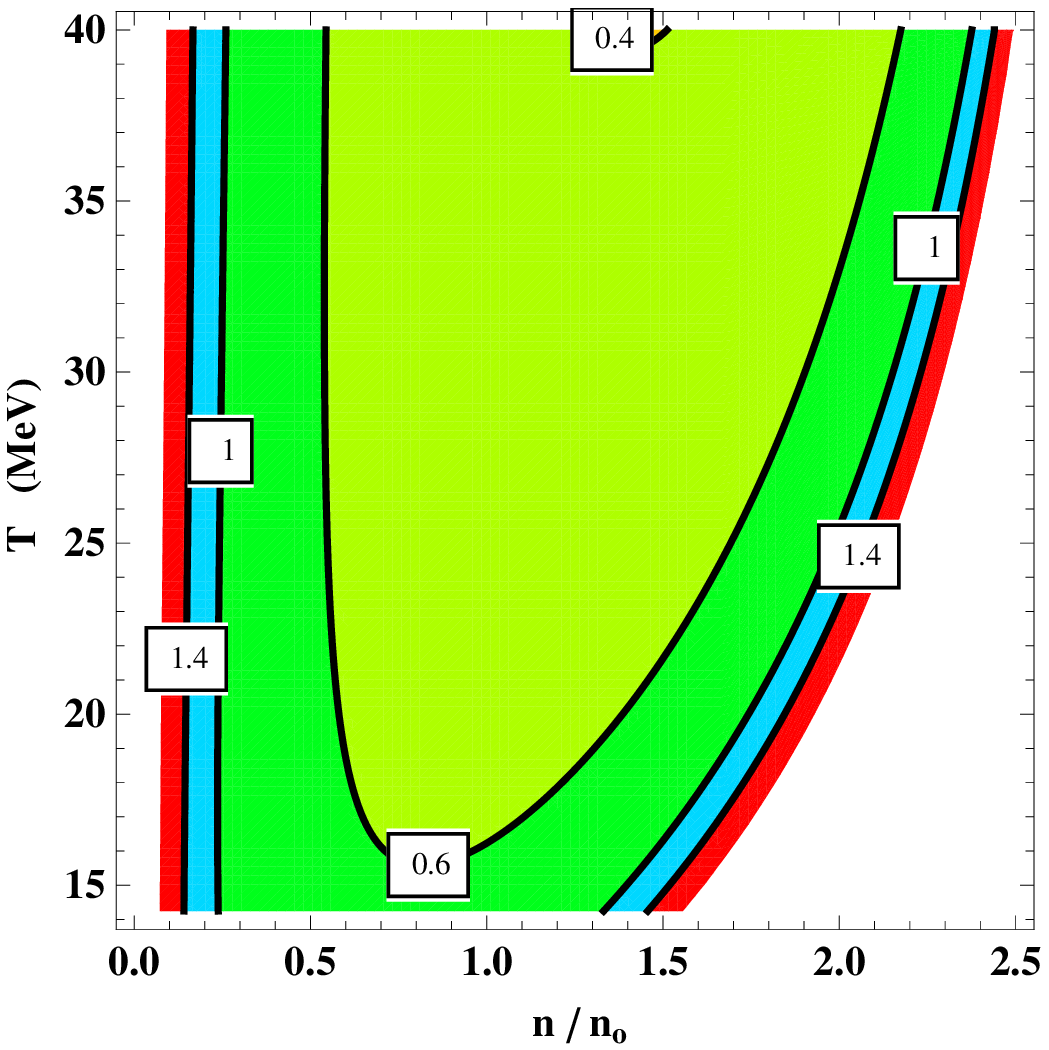}
\end{center}

\vspace{-0.7cm}
\caption{{\small
Same as in Fig.\ \ref{fig3} but for the ratio of the hydrodynamic
CE shear viscosity
[Eq.\ (\ref{etaCE})] to the entropy density
[Eq.\ (\ref{s})].
 }}\label{fig4}
\end{figure*}
Note that the perturbation method for the eigenfunctions
$\varphi(\hat{p})$
(or $\varphi^{}_\ell$, Eq. (\ref{varphipertexp})
as in Ref.\ \cite{chapman})), and in addition,
eigenvalues $\omega$
allows us to obtain
in a regular way high-order corrections
in $\epsilon$. In this way, one
has to go beyond the quadrupole
multipolarity ($\ell \leq 2$) approximation taking into account, consistently
at a given $\epsilon$, higher order terms,  $\ell > 2$, in expansion
(\ref{varphiexp}) for
$\varphi(\hat{p})$. 

\vspace{1.0cm}
\section{DISCUSSION OF THE RESULTS}\label{sec-disc}

Equation (\ref{viscSDfc}) for $\eta$
has the same classical hydrodynamical
dependence on the
temperature $T$ and diameter $d$,
$\eta \propto \sqrt{mT}/d^2$ [cf.\  with Eq.\ (\ref{etaCE})],
because of using the
FC approximation as in both
the molecular kinetic theory \cite{LLv10}
and the CE approach \cite{chapman}.
In this approximation for the overdamped case
(Appendix B)
the dominating contribution into the viscosity
yields from the collision term which mainly determines
both the classical hydrodynamical solutions (Ref.\ \cite{chapman}) 
and our DPW ones [Eq.\ (\ref{planewavesol})] for the
distribution function
to the BVKE. Therefore, as expected, in the limit
$a \rightarrow 0$ ($\mathcal{F} \ll 1$), one finds
the number constant [in front of $\sqrt{mT}/d^2$; see
Fig.\ \ref{fig2} and Eq.\  (\ref{viscSDfc})]
 that approximately
coincides
within the accuracy of  
2\%
with the CE result
(\ref{etaCE}).
The difference between
the hydrodynamical (\ref{etaCE}) and
overdamped DPW (\ref{viscSDfc})
viscosities in the zero interaction constant limit
should be, indeed, small as compared to
the leading collisional term.

Figure \ref{fig2} shows the shear viscosity $\eta$
[Eq.\ (\ref{viscSDfc})]
for a few temperatures above the critical value $T_c$
[Eq.\ (\ref{Tc})]. From Fig.\ \ref{fig2}, one can clearly see that  the shear
viscosity $\eta$ differs significantly
from the classical hydrodynamical formula (\ref{etaCE})
by the particle density dependence. It appears through the
VDW parameter $\mathcal{F}$ [Eq.\ (\ref{Fant})],
owing to accounting for dynamical variations of
the  mean-field interaction (\ref{effpot})
in our derivations. As displayed in this figure,
the significant effects originate by the
Vlasov self-consistent attractive-interaction
terms of the BVKE. In our approach this is achieved by solving
the BVKE (\ref{Boltzlin}) in terms of the DPW 
nonlocal-equilibrium distribution function
$\delta f$ [see Eq.\ (\ref{planewavesol})] and using, therefore,
the perturbation
expansion (\ref{ompertexp}) for the frequency $\omega$ as
a solution of the dispersion equation, in addition to
Eq.\ (\ref{varphipertexp}). 
This is in contrast to the CE approach based on the
dynamical local-equilibrium distribution-function variations 
and hydrodynamical equations,
used on the left-hand side of the BVKE.
The interaction term of the Boltzmann kinetic equation
containing  $\delta U$ [Eqs.\ (\ref{effpot})]
is neglected in the CE method \cite{chapman}
as compared to the integral collision term of the BVKE
at the leading first order in $\epsilon$. Therefore, there is no
particle density corrections to the shear viscosity in the CE approach
at this order. To obtain
these corrections, we found another alternative DPW solution
(\ref{planewavesol}) through the self-consistent interaction term
of the BVKE.
 Note also that with increasing attractive interaction
parameter $a$ ($a>0$), one finds a linearly decreasing viscosity $\eta$
through the dimensionless parameter $\mathcal{F}$.

 Figures  \ref{fig3} and \ref{fig4} show the ratio, $\eta/s$,
of the viscosity $\eta$
to the entropy density $s$
 [Eq.\ (\ref{s})] given by Eqs.\ (\ref{viscSDfc})
and  (\ref{etaCE}), respectively, in the $n-T$ plane for temperatures
$T$ above the critical value $T_c$ [Eq.\ (\ref{Tc})].
As seen from comparison of the two overdamped viscosities
in units of the entropy density in these figures,
the ratio $\eta/s$ takes form of a minimum with
values $\eta/s \siml 1$
at densities $(0.5 - 2)n^{}_0$,
somewhere in a vicinity of the critical point ($T_c$,$n_c$).
 This minimum is significantly smaller
and moves to smaller temperatures in our
 DPW calculations (Fig.\ \ref{fig3}) as compared to
the CE ones (Fig.\ \ref{fig4})  though they are both smaller than 1.
Note also a weak sensitivity of these properties
depending of the size of the hard spheres $d$
around $d=1$ fm for its deflections in about 20\%. However,
$\eta/s\gg 1$ both at small ($n \ll n_0^{}$) and large ($n \simg 2n^{}_0$)
particle density, which makes the ideal hydrodynamic approach
inapplicable for these densities. 
We should emphasize that the BVKE can be
applied for enough dilute system of particles where the mean free path
is large as compared to particles' interaction region (in our example,
of the order of the size of particles $d$).
This gas condition should be satisfied for all desired densities.

\section{CONCLUSIONS}\label{sec-concl}
The shear viscosity of a nucleon gas is derived
by solving the BVKE for the 
FC regime with
taking into account the van der Waals interaction parameters for both the
hard-elastic sphere scattering
and attractive mean-field interaction. The viscosity $\eta$ depends
on the particle density $n$ through the dynamical mean-field forces
measured by the VDW parameter, $~an/T~$, which is positive
for the attractive long-distance mean-field interaction.
Therefore, the viscosity $\eta$
decreases with the interaction constant $a >0$ through the
VDW parameter $\mathcal{F}$. The ratio of the 
FC viscosity to the entropy density, $\eta/s$,  as function of the
particle density $n$ and temperature $T$ is found  to have
a minimum
which is essentially smaller than one. The viscosity is
significantly  smaller at this minimum which moves to smaller
temperatures toward the critical temperature owing to the long-distance
interaction, as compared to the
classical hydrodynamical CE result.  Our
DPW viscosity calculations have
the same overdamped behavior (strong attenuation)
such that  the collisional term is
dominating above all of other
parts of the BVKE.
Note that the viscosity coefficient can be consider as a response
(Ref.\ \cite{balescu,review,hofmann,review2}) of the stress tensor
$\sigma_{\mu\nu}$ for the shear pressure
on the velocity derivative tensor $\mathcal{U}_{\mu\nu}$;
see, e.g., Eq.\ (\ref{etadef}). See also
the Green's--Kubo formula for the
shear viscosity, as for the conductivity coefficient
\cite{balescu,wiranata-jp-14}.

Our results might be interesting for the kinetic and hydrodynamic
studies of nucleus-nucleus collisions at
laboratory  energies of a few hundreds MeV per nucleon.
The ideal hydrodynamics can be a fairly good approximation
for a system of the
interacting nucleons in the region of
$n$ and $T$
 that corresponds at least to $\eta/s\ll 1$.
 However, the classical hydrodynamical
approach for both the dilute nucleon gas
with $n\ll n_0$ and the nuclear-dense matter with $n \simg 2n_0$
seems to be rather questionable to use.
As a different perturbation theory has to be used in
expansions over small $\omega \tau$ for the FC and small
$1/(\omega \tau)$ in the RC regime,
we should expect very different dependencies of the viscosity
(and other transport coefficients) on the particle density $n$
in these two opposite limits. For instance, the 
RC regime is important to study a weak absorption of the DPW in the gas
system with small far-acting interactions, especially for ultrasonic
absorption \cite{bhatia,spiegel_Ultrasonic_2001}.
Therefore, in the case when the contributions of collisions into the BVKE
dynamics are changed from the
dominant (small $\epsilon $) to almost collisionless process
(small $1/\epsilon$) with increasing DPW frequency for a given collision
frequency $\nu$,  a transition from the FC to RC regimes
 should be accounted
beyond the classical hydrodynamical approach.
This can be realized for small $n/n^{}_0$ and
large $\eta/s$,
in the corresponding $n-T$
regions of the phase diagram for analysis of the nucleus-nucleus
 collisions.  Our approach can be applied
to calculations of the thermal conductivity and diffusion
coefficients in nuclear physics, as well as those and viscosity
in nuclear astrophysics, and
to study different phenomena in the electron-ion plasma.

\bigskip
\centerline{{\bf ACKNOWLEDGMENTS}}
\ms
We thank  D.V.\ Anchishkin, S.N.\ Reznik and A.I.\ Sanzhur for
fruitful discussions. The work of M.I.G. was supported
by the Program of Fundamental Research of the Department of
Physics and Astronomy of National Academy of Sciences of Ukraine.
One of us (A.G.M.) is
very grateful for the financial support of the Program of Fundamental
Research  to develop further cooperation with
CERN and JINR ``Nuclear matter in extreme conditions''
by the Department of Nuclear Physics and Energy of National
Academy of Sciences of Ukraine, Grant No. CO-2-14/2016, 
for nice hospitality during his working visit
to the Nagoya Institute of Technology, and
also for financial support from 
the Japanese Society of Promotion of Sciences, Grant No. S-14130.
\medskip

\appendix

\setcounter{equation}{0}

\renewcommand{\theequation}{A.\arabic{equation}}
\renewcommand{\thesubsection}{A\arabic{subsection}}

\section{COLLISION TERM CALCULATIONS}
\l{appA}

We shall neglect approximately an
influence of the effective potential
$\delta U[n(\r,t)]$ (\ref{effpot}) of the long-range
particle interaction
during a two-particle collision  of hard-core sphere particles
of the gas in the FC regime.
Using the multipole expansion (\ref{varphiexp}) of the amplitude factor
$\varphi(\hat{p})$,
one can simplify
the linearized collisional term, $\delta St~$ [Eqs.\ (\ref{dstdef}) and
(\ref{planewavesol})],
in the BVKE (\ref{Boltzlin}),
\bea\l{dst}
&&\!\!\delta St\!=\! \frac{d^2}{4m} \sum_{\ell} \chi^{}_{\ell}
\int \d \p_1\; |\p_1-\p| \int \d \Om_{p'}
 \nonumber\\
&\times&\left\{f^{}_{0}(p_1')f^{}_{0}(p')\left[
Y_{\ell 0}(\widehat{p}_1') + Y_{\ell 0}(\widehat{p}')
\right]\right.\nonumber\\
&-& \left. f_{0}(p_1)f_{0}(p)\left[
Y_{\ell 0}(\widehat{p}_1) +  Y_{\ell 0}(\widehat{p})
\right] \right\}\;,
\eea
where $f^{}_{0}(p)$ is the
static distribution function
(\ref{maxwell}),
\be\l{ftildel}
\chi^{}_{\ell}=\varphi^{}_\ell\;
\exp\left(-i \omega t +i\k\r\right),
\ee
and $\varphi^{}_\ell$ is the $\ell$ coefficient in the expansion 
(\ref{varphiexp})
for amplitudes $\varphi(\hat{p})$.
One finds the relationship between the
impact parameter $\b$ in the
center-of-mass coordinate system [see Fig.\ \ref{fig1}] and the
scattering angle $\theta_{p'}$ (and $ \beta \d \beta$
to $\d\Omega_{p'}$),
\bea\l{impactpar}
\b&=&\cos(\theta_{p'}/2) d,\quad  \b\d \b=\nonumber\\
 &=& \frac{d^2}{8 \pi} \d \Omega_{p'}=
\frac{d^2}{8 \pi} \sin\theta_{p'} \d \theta_{p'} \d \varphi_{p'}\;.
\eea

The Boltzmann collision term (\ref{dst}) is defined
in such a way that its zero and first $\p$ moments have to be zero
because of the particle-number conservation
(related to the continuity
equation),  and momentum conservation,
\be\l{momcons}
 \p + \p_1 = \p' + \p'_1\;,
\ee
(associated with the momentum continuity
equation)
during a two-body collision.
We take also into account that the distribution function
(\ref{maxwell}) is  located within
a small momentum interval $(2mT)^{1/2}$.
Within this range
the momentum vectors
are approximately changed  only by their direction angles,
\be\l{momconsang}
\hat{p} + \hat{p}_1 \approx \hat{p}' + \hat{p}'_1\;,
\ee
and one can use also the
kinetic-energy
conservation equation,
\be\l{conservkinen}
p_{}^{2} + p_1^{2} =
p_{}^{\prime\; 2} + p_1^{\prime\; 2}\;.
\ee
Substituting Eq.\ (\ref{maxwell})
for the static  distribution function,
$f^{}_{0}(p)$, and using
 the conservation
equations (\ref{momcons})--(\ref{conservkinen}) in Eq.\
 (\ref{dst}), one finds
$f^{}_{0}(p_1')f^{}_{0}(p')=f^{}_{0}(p_1)f^{}_{0}(p)$.
Therefore,
from Eq.\ (\ref{dst}), one obtains
\bea\l{dst1}
&&\!\delta St\!=\!\frac{d^2}{4m}\;f_{0}(p)
\sum_\ell\chi^{}_\ell
 \!\int \d \p_1\; |\p_1-\p|\; f_{0}(p_1) \nonumber\\
&\times&\!\int\! \d \Om_{p'}
\left[
Y_{\ell 0}(\hat{p}_1')
\!+\! Y_{\ell 0}(\hat{p}')
\!-\!
Y_{\ell 0}(\hat{p}_1) \!-\! Y_{\ell 0}(\hat{p})
\right].
\eea
Thus, the collisional term (\ref{dst}) ensures all
necessary (particle number, momentum, and energy) conservation laws.
In particular, one can check that
there is no $\ell=0$ and $1$ terms in the sum over $\ell$ of Eq.\
 (\ref{dst1}). By that reason, because of zero two first
moments of the collision term $\delta St$ (\ref{dst1}),
there are no contributions from
(\ref{dst1}) into the continuity equation [zero $\p$ moment of the Boltzmann
equation (\ref{Boltzlin})] and, explicitly,
into the momentum  equation [the first $\p$
moment of (\ref{Boltzlin})]. This term $\delta St$ will affect only on the
momentum flux tensor $\delta \Pi_{\mu\nu}$  (\ref{dpidef}) through the
solutions (\ref{planewavesol}) for the distribution function
$\delta f$ [see, e.g., Eq.\ (\ref{dsigmadef})] in terms of the viscosity
coefficients [Eq.\ (\ref{etadef})].

For the integration over $\p_1$ in Eq.\ (\ref{dst1}),
it is convenient to use the system of the center of mass
for a given two-body collision,
with the symmetry $z$ axis
directed along the relative motion
of projectile  particle having the reduced mass (Fig.\ \ref{fig1}).
We  transform the integral over $\p_1$
to the relative momentum $\q=\p_1-\p$.  Then,
using the spherical-coordinate system
 with the symmetry $z$ axis
directed along the vector $\q$ of the relative motion
of projectile  particle (Fig.\ \ref{fig1}),
$\d \q=q^2 \d q\; \sin \theta_q\; \d \theta_q\; \d \varphi_q$ and
($x_q=\cos\theta_q$), one finds from (\ref{dst1})
\bea\l{dst2}
&&\!\!\delta St^{}_L \equiv
\int \d \Omega_p Y_{L 0}(\hat{p})\delta St(\p)\nonumber\\
&=&\frac{\pi^2 d^2}{4m}\;f_{0}(p)f_{0}(p)
\sum_\ell \chi^{}_\ell\sqrt{(2L+1)(2\ell+1)}\nonumber\\
 \!\!&\times&\!\int_{-1}^{1}\d x P_L(x)
\int_{-1}^{1} \!\d x_q\int_0^\infty q^3\d q
\exp\left[\!-\frac{q^2 \!+\!2 p q x_q}{2 m T}\right]\nonumber\\
\!\!&\times&\!\int_{-1}^{1} \d x'\!
\left[ P_\ell(x_1') \!+\! P_\ell(x') \!-\! P_\ell(x_1) \!-\! P_\ell(x)\right].
\eea
Here,
$P_\ell(\cos\theta_p)=(4 \pi/(2 \ell+1))^{1/2}\;Y_{\ell 0}(\hat{p})$
is the Legendre polynomial of $\ell$ order, and
several transformations of the angle coordinates are performed:
\bea\l{xdefs}
x&=&\cos\theta_{p}\qquad x_1=\cos\theta_{p_1}\;,\nonumber\\
x'&=&\cos\theta_{p'}\qquad x_1'=\cos\theta_{p_1'}\;.
\eea
Note that the integration over the azimuthal angle of the
relative momentum
was taken from zero to $\pi$ \cite{abrkhal}.
For the integration over the modulus of the relative momentum $q$,
for the fixed $x$ and $x'$, one can
change  the angle
variables to functions of the relative $x_q$:
\bea\l{transangles}
x_1'&=&x_1 + x - x'\;,\nonumber\\
x_1&=&z \sqrt{2mT}x_q/p +x\;.
\eea
For the fixed $x$ and $x'$ we integrate first analytically over the variable
$z=q/p^{}_{T}$,  where $p^{}_{T}=\sqrt{2 m T}$, and then
 over $x_q$.
\begin{figure}
\begin{center}
\includegraphics[width=0.49\textwidth,clip=true]{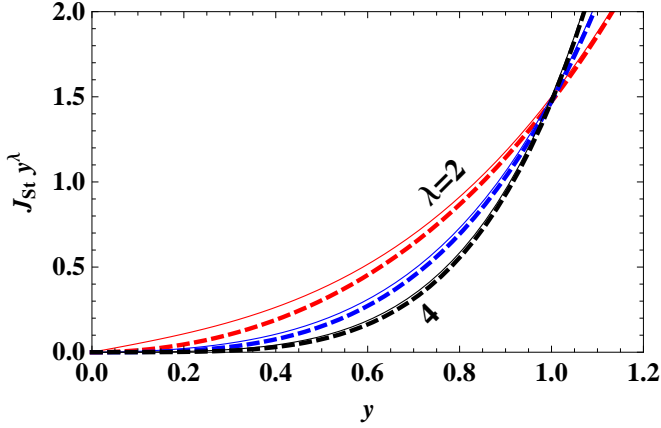}
\end{center}

\vspace{-0.7cm}
\caption{\small
Integral $y^\lambda\mathcal{J}_{St}(y)$ [Eq.\ (\ref{Istfun}),
solid]
 vs $y$ for powers $\lambda=2$ (red), $3$ (blue), and $4$ (black); dashed
lines show those with the corresponding asymptotics
[Eq.\ (\ref{Istfunexplargey})] up to fourth order.
}
 \label{fig5}
\end{figure}
Integrating then, e.g.,  the $\ell=2$ term, $\delta St_2$  of 
Eq.\ (\ref{dst2}), explicitly
 over remaining angles
$x$ and $x'$, one obtains
\be\l{integration}
\delta St^{}_2 =
\frac{5\pi^2 d^2}{2m}p^{4}_{T}\chi^{}_2
f_{0}(p) I_{\rm St}(p),
\ee
where
\bea\l{intSt}
I_{\rm St}(p)&=& f^{}_{0}(p)\int_{-1}^{1} \d x\; P_2(x) \int_{-1}^{1} \d x'
\int_{-1}^{1}\d x_q \nonumber\\
&\times& \int_{0}^{\infty} z^3 \d z\;
\exp\left[- \left(z^2 + 2 p\; x_q z/p^{}_{T}\right)\right]\nonumber\\
&\times&\left[P_2\left(z \sqrt{2mT}x_q/p +2 x - x'\right) +
P_2\left(x'\right)\right.\nonumber\\
&-& \left.P_2\left(z \sqrt{2mT}\;x_q/p + x\right) -
P_2\left(x\right)\right]\nonumber\\
&=& \frac{2 n (2 p^2 + p^2_{T})}{5 \pi\; p\; p^4_{T}}\;
{\rm erf}\left(\frac{p}{p^{}_{T}}\right)
+\frac{4 n}{5\pi^{3/2} p^3_{T}} \nonumber\\
&\times&
\exp\left(-\frac{p^2}{p^2_{T}}\right)
=\frac{4 n}{5 \pi p^3_{T}} \mathcal{J}_{\rm St}(p/p^{}_{T})\;,
\eea
with  the error function
$\mbox{erf}(y)=2 \int_0^y \d z \exp(-z^2)/\sqrt{\pi}$,
\bea\l{Istfun}
&\mathcal{J}_{\rm St}(y)=\frac{y^2+1/2}{y}\;\erf(y)\nonumber\\
&+\frac{1}{\sqrt{\pi}}\;
\exp\left(-y^2\right)\;,\quad
y=p/p^{}_{T}\;.
\eea
To reduce the BVKE to the perturbation eigenvalue
problem [Eqs.\ (\ref{varphipertexp}) and (\ref{ompertexp})] for the
eigenfunctions
$\varphi(\hat{p})$ and eigenvalues $c=\om/(k v^{}_{T})$
as solutions of the linear homogeneous equations for $\varphi(\hat{p})$,
and dispersion equation for $c$ (Appendix B), we may derive now the accurate
constant (independent of $y$) approximations
to the function $\mathcal{J}_{\rm St}(y)$ [Eqs.\ (\ref{Istfun})].
Using these approximations, one obtains Eqs.\ (\ref{tauapprox})
with (\ref{nu}) for the collision term $\delta St$ (\ref{dst2}).
Indeed, we may note that for the derivation of such
approximations the collision term $St$
[Eq.\ (\ref{dst1})] can be considered through all of its $\p$ moments.
They are integrals over the modulus $p$,
which are taken up to the constant from the product of
$\mathcal{J}_{\rm St}(p)$ [Eq.\ \ref{Istfun}]
and the power $p^\lambda$ at $\lambda\geq 2$, in addition to the Maxwell
distribution function $f_{0}(p)$,
\be\l{momiintcol}
\!\!\int_0^{\infty} \d p \;p^\lambda \delta St^{}_2 \propto 
\chi^{}_2\!
\int_0^{\infty}\! \d y \;y^\lambda\mathcal{J}_{\rm St}(y)\;f_{0}(y p^{}_{T}).
\ee
Figure \ref{fig5} shows a fast convergence of the product
$y^\lambda\;\mathcal{J}_{\rm St}(y)$ [Eq.\ (\ref{Istfun})]
of the integrand in
Eq.\ (\ref{momiintcol}) to its asymptotics  at large $y$
in powers of $1/y$ taking
enough many terms,
\be\l{Istfunexplargey}
\mathcal{J}_{\rm St}(y)= y + \frac{1}{2y} +
\mathcal{O}(\frac{1}{y^4}) \;, \quad y \gg 1\;,
\ee
up to fourth-order terms for all $y$ values
owing to the power factor $p^\lambda$ ($\lambda \geq 2$).
Evaluating a smooth asymptotical
function $\mathcal{J}_{\rm St}(y)$ (\ref{Istfunexplargey}) with respect to
the Maxwell distribution function $f^{}_{0}(y p^{}_{T})$ at
the maximum contribution into the integrals (\ref{momiintcol})
at $y \approx 1$
($p \approx p^{}_{T}$),
one obtains approximately the damping rate $\nu$ of the collisional term
(\ref{nu}):
\be\l{nu1}
\nu =
n v^{}_{T}\sigma\;\mathcal{J}_{\rm St}(1) \approx
\frac{3 \pi n v^{}_{T} d^2}{2}\;.
\ee
 Note that the second exponent term in Eq.\ (\ref{Istfun}) for
$\mathcal{J}_{\rm St}$ was exactly
canceled by the second term of the error function expansion, that leads
to a good relative accuracy
(about 6\%) after neglecting terms of the order of $1/y^4$ in asymptotics
(\ref{Istfunexplargey}).

This accuracy can be checked by comparison
of (\ref{nu1}) with calculations of the exact
function $\mathcal{J}_{\rm St}(y)$, and its average
$\langle \mathcal{J}_{\rm St}(y)\rangle_{\rm av}$
over $y$ with the static distribution
$f^{}_{0}$ (\ref{maxwell}),
\be\l{Istfunav}
\hspace{-0.3cm}\langle\mathcal{J}_{\rm St}(y)\rangle_{\rm av}=
\frac{\int_0^\infty y^2\d y \mathcal{J}_{\rm St}(y) f^{}_{0}(y)}{
\int_0^\infty y^2\d y f^{}_{0}(y)}
= \sqrt{\frac{8}{\pi}}.
\ee
Calculating $\nu$ traditionally
\cite{abrkhal,pethsmith2002,smith2005}
through the averaged value (\ref{Istfunav})
of the collision term [ or $\mathcal{J}_{\rm St}(y)$ (\ref{Istfun})]
over all momenta $\p$ (or $y$), one obtains
\bea\l{nuav}
\nu &\approx& \langle \nu\rangle_{\rm av}=
n v^{}_{T}\sigma\;\langle \mathcal{J}_{\rm St}(y)\rangle_{\rm av}
\nonumber\\
&\approx& \sqrt{8/\pi}\;\pi\; n v^{}_{T} d^2\;.
\eea
Thus, both approximations for $\nu$, Eqs.\ (\ref{nu}) and
(\ref{nuav}), are almost the same within a
good relative precision mentioned above.

\setcounter{equation}{0}

\renewcommand{\theequation}{B.\arabic{equation}}
\renewcommand{\thesubsection}{B\arabic{subsection}}
\section{DERIVATIONS OF DISPERSION EQUATION}
\l{appB}

To derive
the dispersion equation (\ref{dispeq}) for the ratio
$c=\omega/(k v^{}_{T})$ with respect to $c$ in the FC regime,
one may specify a small perturbation
parameter $\epsilon$ [Eq.\ (\ref{eps})] in perturbation expansion
for
$\varphi(\hat{p})$ [Eqs.\ (\ref{varphipertexp}) and (\ref{ompertexp})].
Then, in the FC regime
(small $\epsilon $),  one can truncate the expansion
of $\varphi(\hat{p})$ (\ref{varphiexp}) over
spherical functions $Y_{\ell 0}(\hat{p})$
in the  plane-wave distribution
function $\delta f$ (\ref{planewavesol}) at the quadrupole value of $\ell$,
$\ell \leq 2~$, because of a fast convergence of the sum
(\ref{varphiexp}) over $\ell~$ \cite{brooksyk}.
Substituting the plane-wave solution (\ref{planewavesol})
with the multipole expansion
(\ref{varphiexp}) for $\varphi(\hat{p})$
in $\delta f$
into the BVKE (\ref{Boltzlin}),
 after simple algebraic
transformations, one finally arrives
(within the same approximations used in Appendix A)
 at the following linear equations ($L=0,1,2,...$) for $\varphi^{}_\ell$:
\be\l{Boltzeqfin}
\sum_{\ell} \mathcal{A}_{L\ell}(c)\;\varphi^{}_{\ell}=0, \quad
 c=\omega/(k v^{}_{T})\;,
\ee
where
\bea\l{Boltzeqcoef}
&\mathcal{A}^{}_{L\ell}(c) \equiv
c \delta^{}_{L\ell} - C_{\ell 1;L} +
\frac{\mathcal{F}}{\sqrt{3}} \;\delta_{L1}\delta_{\ell 0}\nonumber\\
&+
i \gamma\;\delta_{\ell L}
(1-\delta_{\ell 0})(1-\delta_{\ell 1})\;,
\eea
\bea\l{angintclebsh}
C_{\ell 1;L}&=&\sqrt{\frac{4 \pi}{3}} \; \int \d \Om_p\; Y_{L0}(\hat{p})\;
Y_{10}(\hat{p})\;Y_{\ell 0}(\hat{p})\nonumber\\
&=&\sqrt{\frac{2 \ell+1}{2L+1}}\;
\left(C_{\ell 0,10}^{L0}\right)^2\;,
\eea
$C_{\ell 0,10}^{L0}$ is the Clebsh-Gordan coefficients \cite{varshalovich},
and
$\gamma$ is given by Eq.\ (\ref{gamma})
(Appendix A).
We multiplied the BVKE (\ref{Boltzlin}) by $Y_{L0}(\hat{p})$,
and integrated term by term over angles
$\d\Omega_{p}$ of the unit momentum vector $\hat{p}$ in the spherical
coordinate system with the polar $z$ axis along the unit wave vector $\hat{k}$.
The integrals can be calculated explicitly by using the orthogonal properties
of spherical functions and Clebsh-Gordan techniques for
calculations
of a few spherical function  products in the integrand.
The matrix $\mathcal{A}_{L\ell}$ has a  simple structure.
At the diagonal, one finds non-zero values $\mathcal{A}_{\ell\ell}$
depending on the sound velocity $c$.
There are also two $L=\ell \pm 1$ lines, parallel to the diagonal,
above and below it, with
the nonzero number coefficients, depending on the Clebsh-Gordan coefficients
through Eq.\ (\ref{angintclebsh}).
They are independent of the velocity $c$ and the
dimensionless collisional rate
$\gamma$. Other matrix elements are zero.
The isotropic mean field $\delta U$
(\ref{effpot}) influences, through the interaction constant
$\mathcal{F}$ (\ref{Fant}),
on only one matrix element, $\mathcal{A}_{01}=\mathcal{F}/\sqrt{3}-C_{01,0}$.
The damping rate
constant $\gamma$
related to the collision integral [Eq.\ (\ref{tauapprox})]
 are placed only in the main diagonal
$\mathcal{A}_{\ell \ell}$ at $\ell\geq 2$, 
$\mathcal{A}_{\ell \ell}=c+i \gamma$ because
of the conservation
conditions, as explained above (Appendix A). For the
FC regime, because of large
$\gamma$, one notes the convergence of the
coefficients $\varphi^{}_\ell$ of the expansion in multipolarities
(\ref{varphiexp}): Any
$\varphi^{}_\ell$ at $\ell \geq 2$ is smaller than $\varphi^{}_{\ell-1}$
by factor $1/(c+i \gamma)$
\cite{brooksyk,kolmagpl}. See  more explicit
expressions for
ratios of the amplitudes $\varphi^{}_\ell$
in Appendix C [Eq.\ (\ref{phieq})]
in the case of the quadrupole
truncation of the characteristic matrix $\mathcal{A}~$.
Truncating this matrix
at the quadrupole
value $\ell \leq 2$ and $L \leq 2$, one obtains
the following simple $3\times 3$ matrix
\be\l{matrix2}
\mathcal{A}^{(2)}\!=\!\left(\begin{array}{ccc}
 c &
-C_{11;0} & 0 \\
 \mathcal{F}/\sqrt{3}-C_{01;1} & c & -C_{21;1} \\
 0 & -C_{11;2} & c + i \gamma
\end{array}
\right)\;,
\ee
with $~C_{0 1;1}=C_{1 1;0}=1/\sqrt{3}~$ and
$~C_{2 1;1}= C_{11;2}=2/\sqrt{15}~$. Accounting for
Eq.\ (\ref{gamma}) for $\gamma$,
and explicit expressions for these constants
$C_{\ell 1;L}$ (\ref{angintclebsh}), in the quadrupole 
FC case, one obtains
the condition of
existence  of nonzero solutions [det$\mathcal{A}^{(2)}(c)=0$] of
linear equations (\ref{Boltzeqfin}), that is
the cubic equation (\ref{dispeq}) with respect to $c~$.

Substituting $c=c_r+ic_i$ into the dispersion equation (\ref{dispeq}),
one can use the overdamped conditions within the FC regime,
\be\l{fccond0s}
|c/\gamma|=|\omega \tau| \ll 1, \quad |c_r/c_i|\ll 1\;.
\ee
Then, at leading order one obtains (for $\gamma \neq 0$)
\be\l{eqover}
- i \mathcal{F}_1\;
\frac{c_i}{\gamma} - \mathcal{F}_1\;\frac{c_r}{\gamma}
-i \mathcal{F}_2=0\;,
\ee
where $\mathcal{F}_1=3/5-\mathcal{F}/3$ and
$\mathcal{F}_2=(1-\mathcal{F})/3$,
and  $\gamma$ is given by Eq.\ (\ref{gamma}).
Separating real and imaginary parts, at leading order
within the conditions (\ref{fccond0s}), one finds  the overdamped
solution,
\be\l{sol0soundr}
c_r=0,\quad c_i=-\frac{\mathcal{F}_2}{\mathcal{F}_1}\;\gamma,
\ee
that is identical to Eq.\ (\ref{linsols0}).

\setcounter{equation}{0}

\renewcommand{\theequation}{C.\arabic{equation}}
\renewcommand{\thesubsection}{C\arabic{subsection}}
\section{MOMENTS OF THE DISTRIBUTION FUNCTION AND VISCOSITY}
\l{appC}

For the shear viscosity $\eta$ [Eq.\ (\ref{etadef})], one has to
calculate the matrices
$\mathcal{U}_{\mu\nu}$ [Eq.\ (\ref{veltens})] and
$\delta \sigma_{\mu\nu}$ [Eq.\ (\ref{dsigmadef})]. Taking the
polar axis of the spherical coordinate system in the momentum space
along the unit wave vector $\hat{k}=\k/k$, we note that these
matrices are symmetric with zero nondiagonal terms, and
\be\l{Uxyz}
\mathcal{U}_{xx}=\mathcal{U}_{yy}=-\frac12\;\mathcal{U}_{zz}\;,
\ee
\be\l{sigmaxyz}
\sigma_{xx}=\sigma_{yy}=-\frac12\;\sigma_{zz}\;.
\ee
We find easy these relations using the symmetry arguments
and properties of the integrals of the  plane-wave solution
(\ref{planewavesol}) for $\delta f$ over
the angles $\d \Omega_p$  of vector $\p~$.
Therefore, from Eqs.\ (\ref{veltens}), (\ref{dsigmadef}),
(\ref{planewavesol}) and (\ref{varphiexp})
one has to obtain
only the simplest $zz$ components,
\bea\l{USzz}
\mathcal{U}_{zz}&=&2\frac{\partial u_z}{\partial z}-\frac23 \nabla \u
=\frac43 i k \widetilde{u}_{z}\exp(-i \omega t +i\k\r)\;,
\nonumber\\
\sigma_{zz}&=&\widetilde{\sigma}_{zz}\exp(-i \omega t +i\k\r)\;,
\eea
where
\bea\l{Uzztilde}
\widetilde{u}_{z}&=&
\int \frac{\d \p}{nm} \;p_z f^{}_{0}(p)\;
\varphi(\hat{p})=\frac{v^{}_{T}}{\pi \sqrt{3}}\;\varphi^{}_1\;,\nonumber\\
\varphi_{\ell}&=&\int \d \Omega_p Y_{\ell 0}(\hat{p})\varphi(\hat{p})\;,
\eea
and
\bea\l{Szztilde}
&&\widetilde{\sigma}_{zz}=-\int \frac{\d \p}{3m}\;
\left(3 p_z^2 -p^2\right)\; f^{}_{0}(p)\;\varphi(\hat{p})\nonumber\\
&=& - \frac{2}{3m}\sqrt{\frac{4 \pi}{5}}\int p^4 \d p\; f^{}_{0}(p)
\int \d \Omega_p Y_{20}(\hat{p})\; \varphi(\hat{p})\nonumber\\
&=&-\frac{nT}{\sqrt{5\pi}}\;\varphi^{}_2\;.
\eea
We calculated explicitly the Gaussian-like integrals
over $p$ using the
static distribution function $f^{}_{0}$ [Eq.\ (\ref{maxwell})],
\be\l{intpnu}
\!I_\lambda\!=\!\int_0^\infty \d p p^{\lambda} f^{}_{0}(p)
\!=\!\frac{n p_{T}^{\lambda-2}}{
2\pi^{3/2}}\Gamma\left(\frac{\lambda+1}{2}\right)\;,
\ee
where $\Gamma(x)$ is the $\Gamma$ function.
 Using the orthogonal properties of the
spherical functions
and Eqs.\ (\ref{USzz}), (\ref{Uzztilde}) and (\ref{Szztilde}),
from  Eq.\ (\ref{etadef}), one arrives at
\bea\l{eqdeffin}
\eta= \frac{9i\sqrt{\pi}}{4 \sqrt{15}}\;\frac{nT c}{\omega}\;
\frac{\varphi^{}_2}{\varphi^{}_1}\;.
\eea

So far we did not use a specific regime of collisions and the truncated
linear system
of equations (\ref{Boltzeqfin}).
Solving these equations (\ref{Boltzeqfin}), one obtains
\bea\l{phieq}
\frac{\varphi^{}_0}{\varphi^{}_1}&=& \frac{1}{\sqrt{3}\; c}\;,\nonumber\\
\frac{\varphi^{}_2}{\varphi^{}_1}&=& \frac{2}{\sqrt{15}\;
\left(c + i \gamma\right)}\;.
\eea
With these expressions, from Eq.\ (\ref{eqdeffin}) one obtains
\bea\l{etagenfc}
\eta&=&\frac{3 \sqrt{\pi}}{10}\;\frac{1}{1+c/(i\gamma)}\;\frac{nT}{\nu}
\nonumber\\
&=&\frac{1}{5\sqrt{2\pi}}\;\frac{1}{1+c/(i\gamma)}\;\frac{\sqrt{mT}}{d^2}\;.
\eea
Substituting the overdamped solution for the sound velocity
[Eq.\ (\ref{linsols0})], from Eq.\ (\ref{etagenfc}) one obtains
Eq.\ (\ref{viscSDfc}).

\end{document}